\documentclass[twocolumn,showpacs,prl]{revtex4}

\ifx\pdfoutput\undefined
\usepackage{graphicx}
\else
\usepackage{graphicx}
\fi

\usepackage[center]{subfigure}
\usepackage{bm}
\usepackage{amsmath}
\usepackage{latexsym}

\def\beq{\begin{equation}}
\def\eeq{\end{equation}}

\begin{document}

\title{Optimality Properties of a Proposed Precursor to the Genetic Code}
\author{Thomas Butler and Nigel Goldenfeld}
\affiliation{Department of Physics and Institute for Genomic Biology,
University of Illinois at Urbana Champaign, 1110 West Green Street, Urbana, IL 61801 USA}

\date{\today}

\begin{abstract}

We calculate the optimality of a doublet  precursor to the canonical
genetic code with respect to mitigating the effects of point mutations
and compare our results to corresponding ones for the canonical genetic
code.  We find that the proposed precursor has much less optimality
than that of the canonical code. Our results render unlikely
the notion that the doublet precursor was an intermediate state in the
evolution of the canonical genetic code.  These findings support the
notion that code optimality reflects evolutionary dynamics, and that if
such a doublet code originally had a biochemical significance, it arose
before the emergence of translation.
\end{abstract}


\maketitle

It is now well-established that the canonical genetic code is not a
frozen accident, but exhibits a pattern of amino acid-codon
correspondences that has the effect of making the code insensitive to
certain classes of point mutation or translation
error\cite{SONN65,WOES1965a,ALFF1969,HAIG1991,FREE1998,freeland2003cem,BUTL08}.
A variety of schemes\cite{knight2001rke}, including ones invoking
evolutionary dynamics\cite{VETS2006} and
stereochemistry\cite{knight2003tsg,yarus2005ogc}, have been put forward
to explain this pattern and others\cite{itzkovitz2007gcn} in the
genetic code (for recent reviews, see \cite{Knight_PhD,KOO08}).  It is
important to stress that the optimality of the code is most manifest
with respect to only one particular class of amino acid attributes,
related to the free amino acid polar
requirement\cite{WOES1966a,WOES1966b}, and this suggests the code is a
very ancient part of the cell's machinery, functioning either in its
present role of translation, or in some earlier unknown function.
Additionally, it has been shown recently that the genetic code has
extreme error-minimizing optimality, being more optimal than all but
one or two random codes generated in sets of ten million \cite{BUTL08}.
This result lends strong support to the suggestion that the code's
evolutionary dynamics was dominated by collective mechanisms arising
from horizontal gene transfer\cite{VETS2006}.  Computational evidence
shows that core chemical affinities in the genetic code are fully
compatible with, and independent from, evolutionary dynamics that lead
to error minimizing optimality\cite{CAPO05}, suggesting that
error-minimizing optimality is not a by-product of chemistry but arises
from the evolutionary dynamics.

In this report, we attempt to ascertain to what extent, if any,
error-minimizing optimality can be used to constrain a proposed
scenario for the evolution of the genetic code.  If the optimality with
respect to polar requirement was a feature of the code from very early
times, then precursor code proposals must respect error-minimizing
optimality to a significant degree. Alternatively, proposed precursor
codes may claim to date prior to any code evolution, and to be the
product of other factors alone.  Such precursors would not be expected
to display a significant level of error-minimizing optimality, assuming
that it is indeed the case that optimality is primarily a reflection of
evolutionary dynamics.  Here we show that a specific
biochemically-motivated precursor code does not show evidence for
significant error-minimizing optimality, even though it is a projection
of the canonical code; these results support the notion that
error-minimizing optimality primarily reflects evolutionary dynamics,
and imply that this type of precursor code, if it ever existed, would
have arisen prior to the emergence of translation.

Copley, Smith and Morowitz have suggested that first and, to a lesser
extent, second base assignments in the canonical code would arise if
the code has its origin in amino acid synthesis channels embedded in
dinucleotide complexes prior to the emergence of translation
\cite{COPL05}.  The proposal exploits the strong constraints such a
theory imposes on the first two bases of the genetic code to generate a
specific precursor doublet code based on a projection of the canonical
genetic code to a doublet code.  For most of the projection, the third
codon is sufficiently redundant that the first two bases are sufficient
to define the amino acid coded for by doublet.  In the event that the third
bases associated with a doublet codon code for multiple amino acids, the
proposal favors the simpler of the amino acids (table \ref{1}).  They
further refine the proposal by incorporating possible precursor amino
acids motivated by their study of the biosynthetic pathways for amino
acids (not shown) \cite{COPL05}.

To further assess and characterize the proposed precursor code in
\cite{COPL05} we analyze the degree to which it contains
error-minimizing optimality.  As noted above, the proposed precursor
code is based primarily on arguments about biosynthetic pathways rather
than evolutionary considerations. Additionally, it explicitly dates to
prior to translation \cite{COPL05}.  All mechanisms of which we are aware for
code evolvability explicitly require translation machinery (see for
example \cite{OSAW89, SCHUL94, Knight-Nat_Rev_Genet,KNIG1999,SELL01,
VETS2006}).  Thus we anticipate that the proposed precursor code should
contain little, if any, evidence for optimality.

We have analyzed the former of these proposed precursor doublet codes
(see table) for error-minimizing optimality using the
``experimental polar requirement" (EPR)
\cite{WOES1965a,WOES1965b,WOES1966a,WOES1966b} derived originally by
Woese and co-workers.  We have also analyzed the precursor using a modern
computational update of the polar requirement (CPR) \cite{MATH08}.
Analysis with the CPR is of particular interest, because it is the
measure of amino acid difference that when applied in code optimality
analysis algorithms to the canonical genetic code gives rise to the
extreme optimality cited above \cite{BUTL08}. Thus the CPR can be
considered to capture some essential aspect of amino acid chemistry of
particular relevance during the evolution of the genetic code. Analysis
of the more refined version of the proposed precursor code is difficult
due to the fact that the polar requirements for the proposed precursor
amino acids are unknown.  We believe, however, that the qualitative
results apply to both versions of the proposed precursor since large
changes in the chemical properties as amino acids are refined at the
same position are unlikely.

\begin{table}
    \par
    \mbox{
      \begin{tabular}{|c|c|c|c|c|}
        \hline \multicolumn{5}{|c|}{Proposed Precursor Code}\\ \hline
         & {\bf G}  & {\bf C} & {\bf A} & {\bf U} \\ \hline
        {\bf G} & Gly & Ala & Asp & Val \\ \hline
        {\bf C} & Arg  & Pro & Gln & Leu  \\ \hline
        {\bf A} & Ser  & Thr & Asn & Ile  \\ \hline
        {\bf U} & Cys  & Ser & Tyr & Leu  \\ \hline

      \end{tabular}
      }
      \caption{Proposed precursor code from Ref.\cite{COPL05}.  Row is first base, column is second base.}
      \label{1}
      \end{table}

To analyze the error-minimizing optimality in the proposed precursor
code, we used the point mutation code analysis algorithm described in
\cite{HAIG1991} and \cite{FREE1998}.  The presentation of this
algorithm in \cite{BUTL08} considers an ensemble of random genetic
codes genetic code as mappings from the set of codons (minus the
termination codons) to the set of amino acids, $GC^{i}:Codons
\rightarrow  Amino\ Acids$, where $i$ indexes a particular set of
assignments of codons to amino acids, with $GC^{1}$ as the canonical
code. Versions $GC^{i \neq 1}$ are generated by randomly permuting
amino acid labels, again excluding termination codons.  $O^i$ can then
be calculated as

\begin{equation}
O_{i}^{-1}=\sum_{\langle c,c' \rangle \neq Ter} {(GC^{i}(c)-GC^{i}(c'))^2}
\label{2}
\end{equation}
\noindent where $\langle c,c' \rangle \neq Ter$ denotes a sum over
nearest neighbor codons with the nearest neighbors of a codon defined
by its single point mutations, with all mutations to or from a
termination codon excluded.

To allow simple comparison of the results that does not depend on
rescaling of amino acid properties, we compute the probability
$P_b=Pr(O>O_{1})$ that a random realization is more optimal than the
canonical code by calculating the percentage of our random codes that
are more optimal than the canonical code.

The error in the computed $P_b$ can be estimated using an analytical
realization of bootstrap resampling derived from an exact
correspondence with the statistics of the asymmetric one dimensional
random walk \cite{BUTL08}.  This correspondence shows that if $N$ codes
are sampled, and $N_{O>O_1}$ are more optimal than the code being
tested, then $P_b$ with standard error is given by the expression

\begin{equation}
P_b=(N_{O>O_1}\pm \sqrt{N_{O>O_1}})/N
\label{3}
\end{equation}

While this is in line with naive expectations for the form of error,
the problem of sampling more optimal random codes is a problem of rare
event sampling, which is frequently unstable and prone to nonstandard
large errors.  This makes a rigorous derivation of the exact error a
key result essential for robust interpretation of optimality
calculation results.  The form of the error also informs the
computations.  It is clear from Eq. \ref{3} that the relevant sample
size for a statistically sound analysis is not $N$, but the number of
more optimal codes sampled, $N_{O>O_1}$ \cite{BUTL08}.  A reasonable
minimum is, perhaps, 20 more optimal codes sampled to get a statistical
estimate. Much larger samples would be preferable, but in many
applications may be hard to obtain due to computational limitations
encountered when analyzing highly optimized codes.

When applied to the proposed precursor code, we estimated $P_b=(1.44
\pm 0.038)\times 10^{-2}$ with the experimental polar requirement, or
$P_b=(7.95\pm 0.282)\times 10^{-3}$ with the computational polar
requirement \cite{MATH08}.  To compare, we applied this simplified code
analysis algorithm to the canonical genetic code.  The canonical
genetic code has optimality of $P_b=(1.18\pm 0.109)\times 10^{-4}$ or
$P_b=(4.7\pm 0.686)\times 10^{-5}$ with the EPR and CPR respectively
(the extreme optimality discussed above included transition and
transversion biases for each base position in the calculation
\cite{FREE1998,BUTL08}).  Thus the optimality of the precursor is, with
either the EPR and the CPR, two orders of magnitude less optimal than
the canonical genetic code evaluated with the equivalent algorithm.

As discussed above, the derivation of the doublet code in table \ref{1}
depended on projecting the third base onto the doublets by favoring the
simplest amino acid coded for by the triplet codons associated with a
given doublet.  We repeated the optimality analysis for versions of the
doublet code that favored more complex amino acids at individual
doublets (such as substituting Arg for Ser at the AG position).   None
of the modified doublet codes displayed a significant increase in
optimality over the version in table \ref{1}.  Thus a more optimal
version of the precursor code, which respects the underlying
biosynthesis theory, would differ in several positions from the proposal
by Copley et al.\cite{COPL05}.

Our results show that the proposed precursor code has weak
error-minimizing optimality with respect to the polar requirement,
compared to the canonical genetic code. This result is surprising in
one respect, because the doublet code is a projection of the canonical
code.  A number of possible interpretations are possible.  (1) The
doublet precursor code is not an intermediate evolutionary stage from
some earlier precursor code; this is consistent with the basis for the
original proposal of this code as a biosynthetic pathway, but is
puzzling because the later canonical triplet code is optimized with
respect to the free amino acid polar requirement.  (2) The precursor
has no biological significance at all, and did not evolve from an
earlier precursor, which exhibits free amino acid polar requirement
optimality.  (3) The precursor doublet code predates evolution for
error minimization, and if the amino acid synthesis scheme is correct,
then modifications to the doublet code during its evolution to today's
canonical code are responsible for its observed error-minimizing
optimality.  The relatively large $P_b$ value (i.e. small amount of
observed optimality) in the precursor is an artifact of deriving the
doublet code from the highly evolved canonical code.

Our analysis does not address the question of whether or not the
detailed biochemical theory proposed is correct, because presumably
optimal precursor codes that are consistent with both the biochemical
theory and uncorrupted by evolution could be constructed.

We gratefully acknowledge discussions with Carl Woese, Rob Knight,
Shelley Copley, Eric Smith and Harold Morowitz.  This material is based
on work supported by the National Science Foundation under Grant No.
NSF-EF-0526747.

\bibliographystyle{apsrev}

\bibliography{doublet}

\end{document}